# **Extrinsic *n*-type semiconductor transition in $ZrSe_2$ with the metallic character through hafnium substitution**

Zahir Muhammad[a], Yuliang Li[b], Sami Ullah[c], Firoz Khan[d], Saleh S Alarfaji[e], Abdulaziz M. Alanazi[f], Zhe Sun[b], Thamraa Alshahrani[g,*], Yue Zhang[a,*], Weisheng Zhao[a]

[a]*Hefei Innovation Research Institute, School of Integrated Circuit Science and Engineering, Beihang University, Hefei 230013, PR China*

[b]*National Synchrotron Radiation Laboratory, University of Science and Technology of China, Hefei 230029, China*

[c]*K. A. CARE Energy Research & Innovation Center (ERIC), King Fahd University of Petroleum & Minerals (KFUPM), Dhahran 31261, Saudi Arabia*

[d]*Interdisciplinary Research Center for Renewable Energy and Power Systems (IRC-REPS), King Fahd University of Petroleum & Minerals (KFUPM), Dhahran 31261, Saudi Arabia*

[e]*Department of Chemistry, Faculty of Science, King Khalid University, P.O. Box 9004, Abha 61413, Saudi Arabia*

[f]*Department of Chemistry, Faculty of Science, Islamic University of Madinah, Madinah 42351, Saudi Arabia*

[g]*Department of Physics, College of Science, Princess Nourah bint Abdulrahman University, P.O. Box 84428, Riyadh 11671, Saudi Arabia*

*Email:thmalshahrani@pnu.edu.sa, yz@buaa.edu.cn

## Abstract:

Two-dimensional (2D) layered materials exhibit versatile electronic properties in their different phases. The intrinsic electronic properties of these materials can be modulated through doping or intercalation. In this study, we investigated the electronic properties of Hf-doped $ZrSe_2$ single crystals using angle-resolved photoemission spectroscopy (ARPES) combined with first-principles density functional theory (DFT) calculations. It is observed that the valence band maxima of $ZrSe_2$, located below the Fermi level, undergo a significant change with the introduction of Hf substitution. Hf can introduce extra charges into the conduction band, rather than making a mixed structure of $HfSe_2$ and $ZrSe_2$ band structure, which can cross the Fermi level. Compared to the semiconducting band structure of $ZrSe_2$, we observed that the conduction band crosses the Fermi level at the high symmetry M point in Hf-doped $ZrSe_2$. This suggests an increase of electron-type carriers around the Fermi level, resulting in an extrinsic charge carrier density in the conduction band, which can form a metallic behaviour. It can be noticed that the Hf cations can create disorder in the form of excess atoms of Zr, which yields more carriers in the conduction band in the shape of "smeared bands". The tails of the "smeared band" occupied the *d*-orbitals extended into the Fermi level and left the *d*-band below. Similarly, the electrical resistance measurements further confirm the metallic-like character of Hf-doped $ZrSe_2$ compared to the semiconductor $ZrSe_2$, indicating increased carriers. This metallic-like behavior is suggested to be predisposed by the extrinsic electrons induced by the substitutional disorder. This study further demonstrates the possibility of band gap engineering through heavy metal doping in 2D materials.

## Introduction

The transition metal dichalcogenides (TMDs) exhibit unique and distinct electronic characteristics in different hybrid structures. Changing the band gap property allows tuning the electronic and optical properties [1]. TMDs possess two phases with diverse electronic structures: the semiconductor 2H phase and the metallic 1T phase [2–4]. These phases coexist in two-dimensional (2D) TMDs [5], and their conversion can be achieved through atomic intercalation, atom substitution, or atomic plane gliding [2, 6–9]. Phase transitions in TMDs depend on growth conditions such as temperature and pressure [10]. The transition metal is sandwiched between two atomic chalcogens in a TMD layer, offering additional degrees of freedom in structural transformations [11]. The electron-phonon coupling is attributed to the structural instabilities in metallic layered transition metal dichalcogenides. The phonon-electron instability may somehow be connected to the electron or hole-type carriers, which changes the phase in layered crystals. However, the band structures of the crystals reveal the presence of saddle points near the Fermi surface, indicating the occurrence of a phase transition [12].

Electronic phase transition plays a crucial role in various physical properties [13–15]. TMDs possess versatile electronic structures that find applications in p-n junctions and other applications [16], high-mobility field-effect transistors [17–19], switching and memory devices [20], and optoelectronics [21]. 2D layered TMD materials can also be found in hybrid forms [22, 23] with distinct electronic structures. The electronic properties of TMDs can be modulated through substitution or intercalation [24–26], leading to altered electronic properties and new applications [27–31]. Atomic substitution or intercalation effectively modifies the intrinsic electronic properties by altering the band gap characteristics [32–34]. For instance, heavy atomic doping of W into $MoTe_2$, $MoSe_2$, and $MoS_2$ lattices induces a transition from semiconducting to metallic phases [27, 35–37]. Strain can be used to tune the band gap properties of IVB TMDs (e.g., $TiX_2$, $ZrX_2$, and $HfX_2$, where X = S, Se, and Te) [38]. Intercalation of Cu in $ZrSe_2$, a semiconductor, provides extra electrons to the d-orbitals of zirconium atoms, leading to a metallic transition [39]. Similarly, in $ZrSe_2$, the conduction band maximum (CMB) crosses the Fermi level at the M point due to the in situ surface doping of sodium, acting as an electron dopant [40]. Intercalation of Cu in 2H-$TaSe_2$ [41], Se doping 1T-$ZrTe_2$ [42], and Cu intercalation 2H-$NbSe_2$ [43] can alter the electronic structure by introducing charge carriers. Recently, Yongji *et al.* investigated the semiconductor-to-metallic transition in $SnS_2$ through Co-intercalation, which modifies the electronic structure [9]. The strong interaction between electrons and holes has prompted the proposal of a mechanism to explain the phase transition.

In this study, we examine the electronic properties of $Zr_{1-x}Hf_xSe_2$ (x = 0.12, determined via ICP) single crystal alloys compared to $ZrSe_2$. The samples were synthesized using the chemical vapor transport (CVT) method. The electronic structures were analyzed using synchrotron-based Angle-Resolved Photoemission Spectroscopy (ARPES) and first-principles density functional theory (DFT) calculations. Our findings reveal a substantial change in the band structures of pristine $ZrSe_2$ upon substituting Hf atoms. In $Zr_{0.88}Hf_{0.12}Se_2$, the electronic properties exhibit additional electron carriers near the bottom of the CB at the M point, crossing the Fermi level and behaving as a metallic character. However, the overall structure is intact and preserved.

## Experiments

### Single crystal growth:

The single crystals of $ZrSe_2$ and $Zr_{1-x}Hf_xSe_2$ were synthesized by using the chemical vapor transport (CVT) method. A highly purity of 99.98% zirconium (Zr) and hafnium (Hf) powders from (Alfa-Aesar, UK) and 99.98% Se from (Sigma-Aldrich, USA) was used for the single crystal growth. The stoichiometric amount with a transport agent of iodine in the ratio of 5 mg/ml was ground for some time to mixed very well and kept in a quartz tube. Later, the tubes containing the accumulated amount of the reactants and purified iodine were attached to the vacuum line with argon gas to evacuate and seal silica tubes with pressure at ~$10^{-4}$ torr. The evacuated quartz tubes were than placed into a two-zone furnace with reaction temperature of 950 °C and a growth temperature of 850 °C. The temperature of the furnace was gradually increased to the transport temperature in 5 hours. With the increase in temperature, the vapor pressure was increased, and the reaction of selenium with the Zr and Hf was continued slowly. The ampule was kept at this temperature for 5 days, after that it was slowly cooled down to 500 °C in 5 hours, followed by a transition to room temperature. After finishing the experimental process, we successfully attained the high-quality single crystals of pristine $ZrSe_2$ and $Zr_{1-x}Hf_xSe_2$ (shown in **Fig. 1**). After proper cleaning the samples was further characterized.

## Results and discussion

We synthesized $ZrSe_2$ and Hf-doped $ZrSe_2$ using the CVT method. **Fig. 1(a)** provides a schematic illustration of the CVT process, images of the single crystal inside the quartz tube and the structural configuration of $ZrSe_2$ and Hf-doped $ZrSe_2$. Additional details regarding the growth conditions of the single crystal can be found in the experiment part. **Fig. 1(b)** presents the X-ray diffraction (XRD) patterns of $ZrSe_2$ and Hf-doped $ZrSe_2$. All XRD patterns closely match the standard diffraction pattern of $ZrSe_2$ (JCPDS 65-3376). However, the diffraction peaks of Hf-doped $ZrSe_2$ exhibit slight shifts to higher diffraction angles and a new peak (-103) corresponding to HfSe. These slight peak shifts in the diffraction patterns towards higher angles are attributed to the smaller ionic radii of Hf ($Hf^{4+}$) compared to Zr ($Zr^{4+}$) during the substitution process [44]. XRD analysis further reveals that the c-lattice parameter of $ZrSe_2$ is c = 6.16 Å, which decreases to c = 6.19 Å for Hf-doped $ZrSe_2$. The lattice expansion occurs due to the larger atomic number (*Z*) of hafnium atoms. A similar trend has been previously observed in $Hf_xZr_{1-x}NiSn_{0.99}Sb_{0.01}$ [45].

The energy-dispersive X-ray spectroscopy (EDS) was performed to estimate the appropriate amount of each element in the single crystal of Hf-doped $ZrSe_2$ (see **Fig. S1**, in supporting information); while further confirmed by Inductively coupled plasma mass spectrometry (ICP). From the EDS and ICP, we have confirmed the elemental composition as $Zr_{0.88}Hf_{0.12}Se_2$. Additionally, X-ray photoelectron spectroscopy (XPS) also revealed the detailed elemental composition along with the chemical states of Hf, Zr and Se, respectively.

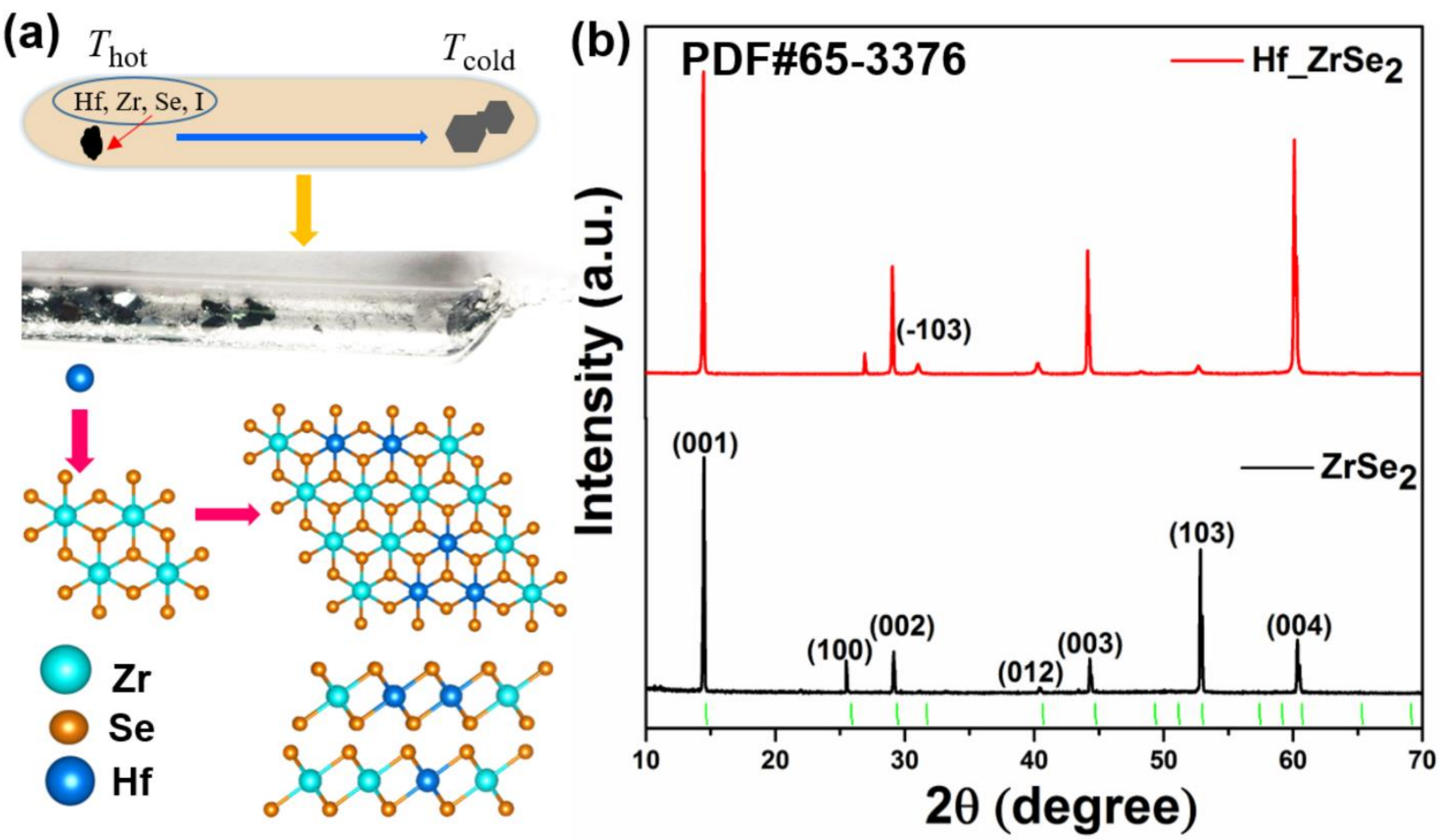

**Figure 1.** (a) Schematic of the crystal growth, image of the crystal and crystal configuration of the $ZrSe_2$ and Hf-doped $ZrSe_2$ (b) X-ray diffraction of $ZrSe_2$ and Hf-doped $ZrSe_2$.

The XPS spectra were analyzed to investigate the chemical states of $Zr_{0.88}Hf_{0.12}Se_2$ compared to $ZrSe_2$, as depicted in **Figs. 2 (a-c)**. **Fig. 2(a)** displays the Zr-3$d_{5/2}$ and Zr-3$d_{3/2}$ peaks. On the other hand, Figure 2(b) illustrates the splitting of the Se-3*d* band in $Zr_{0.88}Hf_{0.12}Se_2$, which is slightly shifted to lower binding energy compared to the $ZrSe_2$ spectra. The shift of the Zr peaks towards higher binding energies after Hf substitution can be attributed to charge transfer, while the Se peaks remain relatively unchanged. This shift has been previously observed in similar studies [46, 47]. The primary cause of this shift is the reduction in bond length or lattice contraction, which increases the binding energy of the Zr 3*d* core level spectra. Similar observations have been made in $Me_xTiSe_2$ (Me = Cr, Mn, Cu) [48] and $V_xTi_{1-x}Se_2$ [47]. **Fig. 2(c)** clearly shows the presence of $Hf^{4+}$ (as indicated by the binding energy position of Hf-4*f*) rather than $Hf^0$, demonstrating the formal $Hf^{4+}$ oxidation state [49]. As the $Hf^{4+}$ substitutes on the $Zr^{4+}$ site, the charge carriers differ from those in the pure sample, resulting in a shift to lower binding energies. Therefore, the extrinsic charge contribution modifies the valence states, leading to a change in the binding energy position.

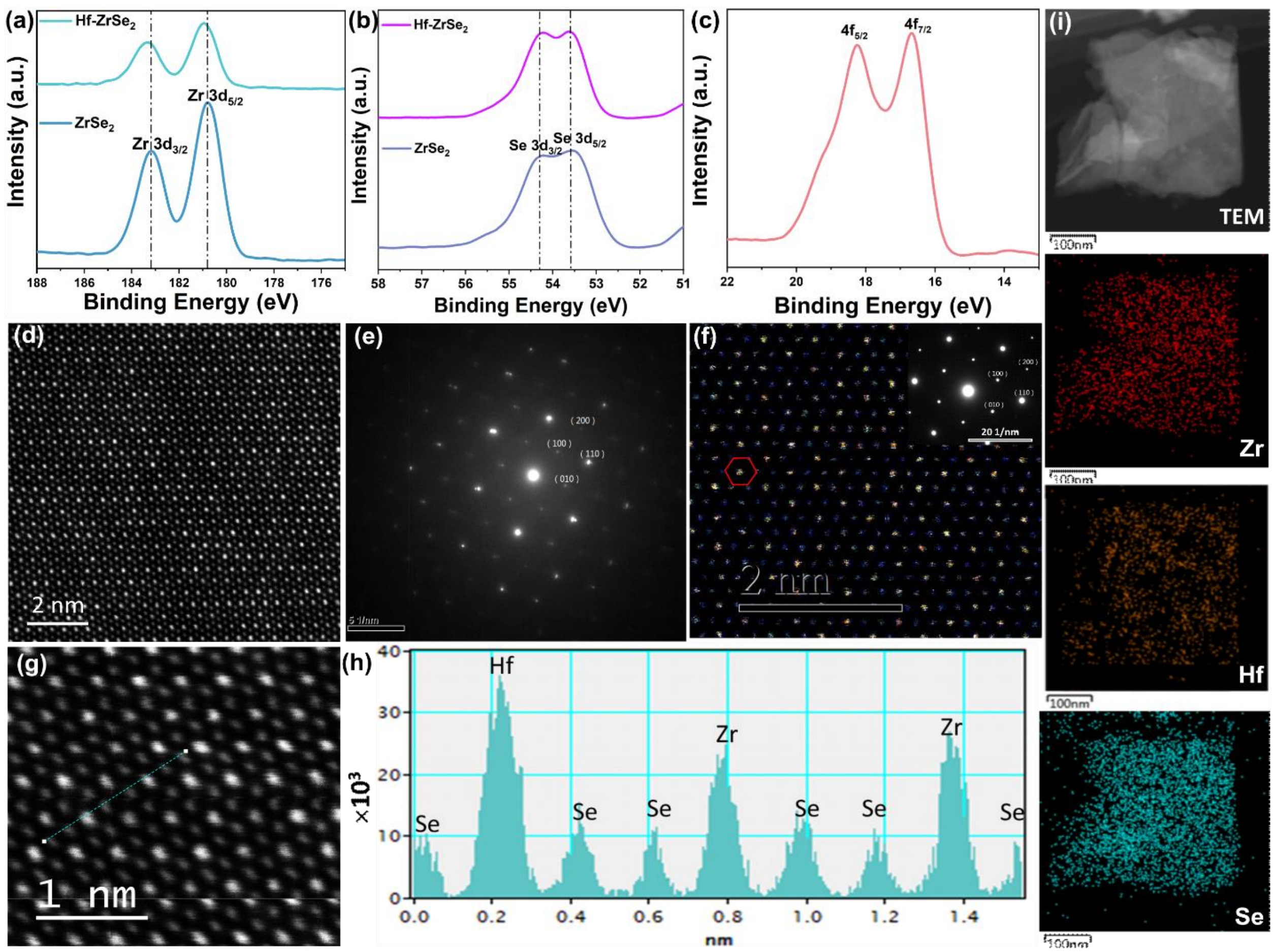

**Figure 2.** Microstructural and elemental analysis of $Zr_{0.88}Hf_{0.12}Se_2$ in comparison with $ZrSe_2$. XPS spectra of (a) Zirconium (Zr 3d), (b) Selenium, (Se 3d) and (c) Hafnium (Hf 4f). (d) Z-contrast high-resolution HAADF STEM images of $Zr_{0.88}Hf_{0.12}Se_2$ and its corresponding (e) SAED patterns, (f) $ZrSe_2$ and inset: SAED patterns. (g) HAADF-STEM image of $Zr_{0.88}Hf_{0.12}Se_2$ and the corresponding (h) intensity profile along the line marked in (g). (i) Typical TEM image of $Zr_{0.88}Hf_{0.12}Se_2$ and corresponding elemental mapping (Zr, Hf and Se, respectively).

Scanning transmission electron microscopy (STEM) was employed to investigate the microstructure of $Zr_{0.88}Hf_{0.12}Se_2$ compared to $ZrSe_2$. **Fig. 2(d)** presents a high-resolution Z-contrast HAADF STEM image of $Zr_{0.88}Hf_{0.12}Se_2$, revealing the hexagonal arrangement of Zr/Hf atoms with six Se atoms surrounding them. **Fig. 2(e)** shows selected-area electron diffraction (SAED) patterns, which exhibit regular diffraction superlattice spots along the rotational direction in the hexagonal axis. Comparatively, the microstructure of $ZrSe_2$, as depicted in **Fig. 2(f)**, displays a similar arrangement. The inset in **Fig. 2(f)** shows SAED patterns of $ZrSe_2$ with similar diffraction patterns. The microstructure of both compounds is similar to each other rather than structural change. In order to reduce noise, a Wiener filter is applied to the high-resolution HAADF-STEM image of $Zr_{0.88}Hf_{0.12}Se_2$, as shown in **Fig. 2(g)**. **Fig. 2(h)** displays the corresponding intensity profile along a line cut in **Fig. 2(g)**, indicating three distinct atomic column profiles. The profile with the highest intensity corresponds to Hf atoms ($Z = 72$), while the less intense profile corresponds to Se atoms ($Z = 32$). Regarding peak intensity, the intensity profile of Zr atoms ($Z = 40$) lies between Hf and Se. This confirms that Hf is substituted on the Zr site in the crystal structure.

Furthermore, **Fig. 2(i)** displays a typical TEM image of $Zr_{0.88}Hf_{0.12}Se_2$ along with the corresponding elemental mapping, revealing a uniform distribution of Zr, Hf, and Se atoms. The interlayer spacing of both samples is confirmed by high-resolution transmission electron microscopy (HRTEM) images (refer to **Fig. S2** in the supporting information). The interlayer spacing of $Zr_{0.88}Hf_{0.12}Se_2$ is observed to be smaller than that of $ZrSe_2$, which is consistent with the X-ray diffraction (XRD) analysis.

The Raman spectra were used to compare the in-plane and out-of-plane structural properties of $Zr_{0.88}Hf_{0.12}Se_2$ with the pristine sample. The lattice vibration is observed to be well retained compared with $ZrSe_2$ (**Fig. S3**, in supporting information), showing that the sample's structure is unchanged. The Raman peaks of $Zr_{0.88}Hf_{0.12}Se_2$ show a minor energy shift (1.9 $cm^{-1}$) for in-plane vibration mode $E_g$ to lower energy, which exhibits the reduction of photon energy. Similarly, the out-of-plane vibration $A_{1g}$ mode shows a slight high energy shift (2.8 $cm^{-1}$) to higher energy, demonstrating the local shortening of Zr-Se covalent bonds due to Hf substitutions. A similar trend was also observed in $Hf_xTi_{1-x}Se_2$ and $Zr_xTi_{1-x}Se_2$ [50, 51]. These findings are further supported by X-ray absorption fine structure (XAFS) measurements.

X-ray absorption near-edge structure (XANES) spectroscopy experiments were conducted to investigate the possible corrugations of $Zr_{0.88}Hf_{0.12}Se_2$ compared with $ZrSe_2$ as presented in **Fig. 3**. **Fig. 3(a)** illustrates the Zr *K*-edge XANES spectra of $Zr_{0.88}Hf_{0.12}Se_2$ in comparison with Zr *K*-edge of $ZrSe_2$ single crystals. The slight chemical shift in energy for $Zr_{0.88}Hf_{0.12}Se_2$ strongly depends upon the charge transfer in the ionic bond and the number of nearest neighbor atoms [52, 53]. Therefore, it has been investigated that the valence and conduction states of the elements are slightly affected in the case of Hf-doping. **Fig. 3(b)** displays the corresponding Fourier transform (FT) spectra, which can show the spectral shape of the atomic structure for both $ZrSe_2$ and $Zr_{0.88}Hf_{0.12}Se_2$. The FT curves indicate two peaks at ~2.31 and 3.45 Å, corresponding to Zr-Se and Zr-Zr correlations of $ZrSe_2$ and $Zr_{0.88}Hf_{0.12}Se_2$, respectively. The intensities of these two peaks are marginally decreased, which suggests that the Zr-neighboring structure of $Zr_{0.88}Hf_{0.12}Se_2$ has changed as associated with $ZrSe_2$. The minor shift shows the decrease of metal-metal bond length from 3.79 to ~3.43 Å (see Table S1).

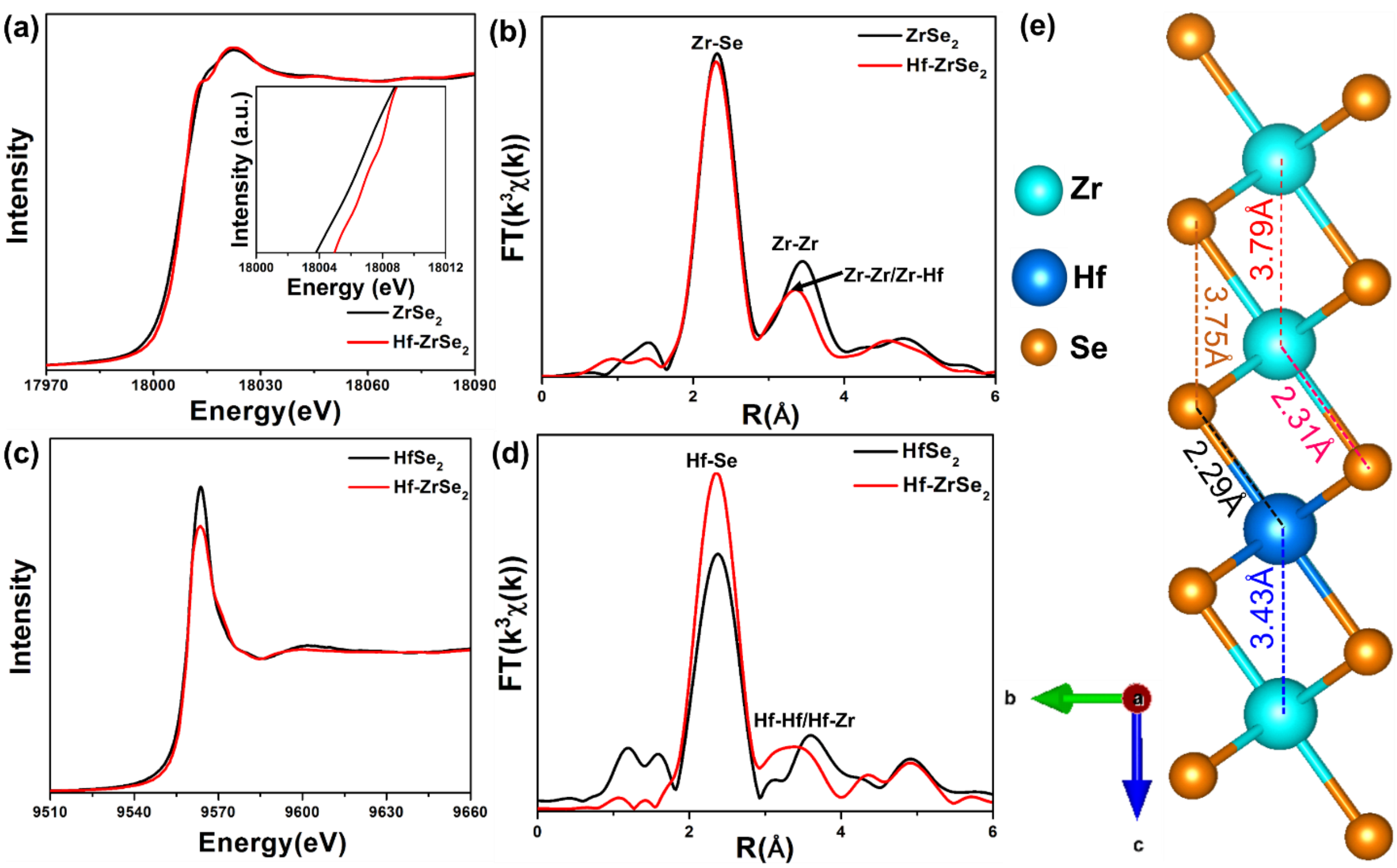

**Figure 3.** XAFS analysis of as-grown $ZrSe_2$ and $Zr_{0.88}Hf_{0.12}Se_2$ single crystals (a) Zr K-edge XANES spectra of $Zr_{0.88}Hf_{0.12}Se_2$ compared with $ZrSe_2$ single crystals and the corresponding (b) Fourier transforms ($FT(k^3\chi(k))$) curves. (c) Hf L3-edge $Zr_{0.88}Hf_{0.12}Se_2$ compared with pure $HfSe_2$ and its equivalent (d) Fourier transforms ($FT(k^3\chi(k))$) spectra. (e) Structural models showing the bond length of pure $ZrSe_2$ and $Zr_{0.88}Hf_{0.12}Se_2$.

Similarly, **Fig. 3(c)** shows the comparison of Hf *L*3-edge XANES spectra of $Zr_{0.88}Hf_{0.12}Se_2$ with the Hf *L*3-edge of $HfSe_2$, which are similar to each other except slight shift to higher energy was observed for $Zr_{0.88}Hf_{0.12}Se_2$ sample. **Fig. 3(c)** revealed that Hf is bonded well with Se. The slight increase in the intensity of the Hf *L*3-edge of $HfSe_2$ compared with $Zr_{0.88}Hf_{0.12}Se_2$ shows the different alternation of dissimilar atoms in $Zr_{0.88}Hf_{0.12}Se_2$. Similarly, **Fig. 3(d)** shows the FT curve of $Zr_{0.88}Hf_{0.12}Se_2$ in association with $HfSe_2$, with a smaller Hf metal-metal bond length shift from 3.79 to ~3.43 Å (see **Table S1**). The minor difference in the structure is the cause of bond length shortening for the doped case. The XAFS results are well in accordance with XRD and Raman data, which can support the extrinsic charge contribution. The XAFS spectra were subjected to further analysis to obtain quantitative structural results. This was achieved by performing least-squares fittings of the Zr and Hf K-edge data, and the summarized findings can be found in **Table S1** of the supporting information. **Fig. 3(e)** shows the structural model based on the structural parameters obtained from XAFS data, which illustrates the different bond lengths of the Zr-Se, Zr-Zr, Zr-Hf, Hf-Se, Zr-Se-Zr, Zr-Se-Hf and Hf-Se-Hf etc. with different coordination (see **Table S1**). Compared with pure $ZrSe_2$, the Zr-Se and Zr-Zr bond length in the $Zr_{0.88}Hf_{0.12}Se_2$ was relatively shortened, thus impairing the Hf to make a covalent bond on the Zr site.

To investigate the electronic band structures of $Zr_{0.88}Hf_{0.12}Se_2$ compared with the parent compound using angle-resolved photoemission spectroscopy (ARPES) for a detailed study. Further details of the ARPES measurement can found in the supporting information. Notably, we know that the nature of $ZrSe_2$ is semiconducting, which is well understood, while it needed to be clarified whether the Hf-doped $ZrSe_2$ ternary alloy is still a semiconducting phase. Therefore, the electronic structure of $Zr_{0.88}Hf_{0.12}Se_2$ single crystals compared to pure $ZrSe_2$ was investigated using ARPES combined with DFT calculations as shown in **Fig. 4**. The details of the DFT calculation can be found in the supporting information.

**Fig. 4(a)** shows the corresponding structure's Brillouin zone (BZ), indicating the high symmetry *k*-points. **Fig. 4(b)** indicates the constant energy contour plots $ZrSe_2$ and $Zr_{0.88}Hf_{0.12}Se_2$ along the high symmetry ΓM in the first BZ. The upper panel in **Fig. 4(b)** shows the Fermi surface (FS) of the pristine $ZrSe_2$ sample, and the lower panel indicate the FS of $Zr_{0.88}Hf_{0.12}Se_2$. It is observed that the spectral weight appeared at the FS along the M direction of the BZ compared to $ZrSe_2$, having no spectral weight on FS. It is suggested that substituting the Hf atom on the Zr site shows excess charges, which can cross the bottom of the CBM at the Fermi level (FL). These tiny electron pockets are visible at the FS at high symmetry M point, which probably comes from the bottom of the *d* band. In principle, the extrinsic electron concentrations due to Hf doping can create large spectra weight of electron pockets in *k* space, which is clearly visible at the FS. In **Fig. 4(c, d)**, we compared the band dispersion of both $ZrSe_2$ (adopted from our previous work [39]) and $Zr_{0.88}Hf_{0.12}Se_2$ along ΓM direction in a large range of *k*-space. By comparing the band structure of $ZrSe_2$ with $Zr_{0.88}Hf_{0.12}Se_2$, it can be observed that the top of the valence band maximum (VBM) of pristine and $Zr_{0.88}Hf_{0.12}Se_2$ are located at about -0.87 eV and -1.05 eV below the Fermi level, respectively, suggesting a non-linear variation of the band gap with Hf concentration due to charging. With Hf substitution, the VBM moves approximately 0.18 eV to deeper binding energy without any evident changes in the band dispersion. Starnberg *et al.* observed the same effect in the photoemission study of $Hf_xTi_{1-x}Se_2$ alloy [54]. Meanwhile, the Fermi level leaves down the bottom of the CB around the zone boundary marked by the red circle and **Fig. 4(d)** and can be clearly seen from their second derivative. These results indicate the extrinsic electron concentration in $Zr_{0.88}Hf_{0.12}Se_2$. The sample shows extrinsic *n*-type behavior of the metallic character, which can be further observed from resistivity measurement.

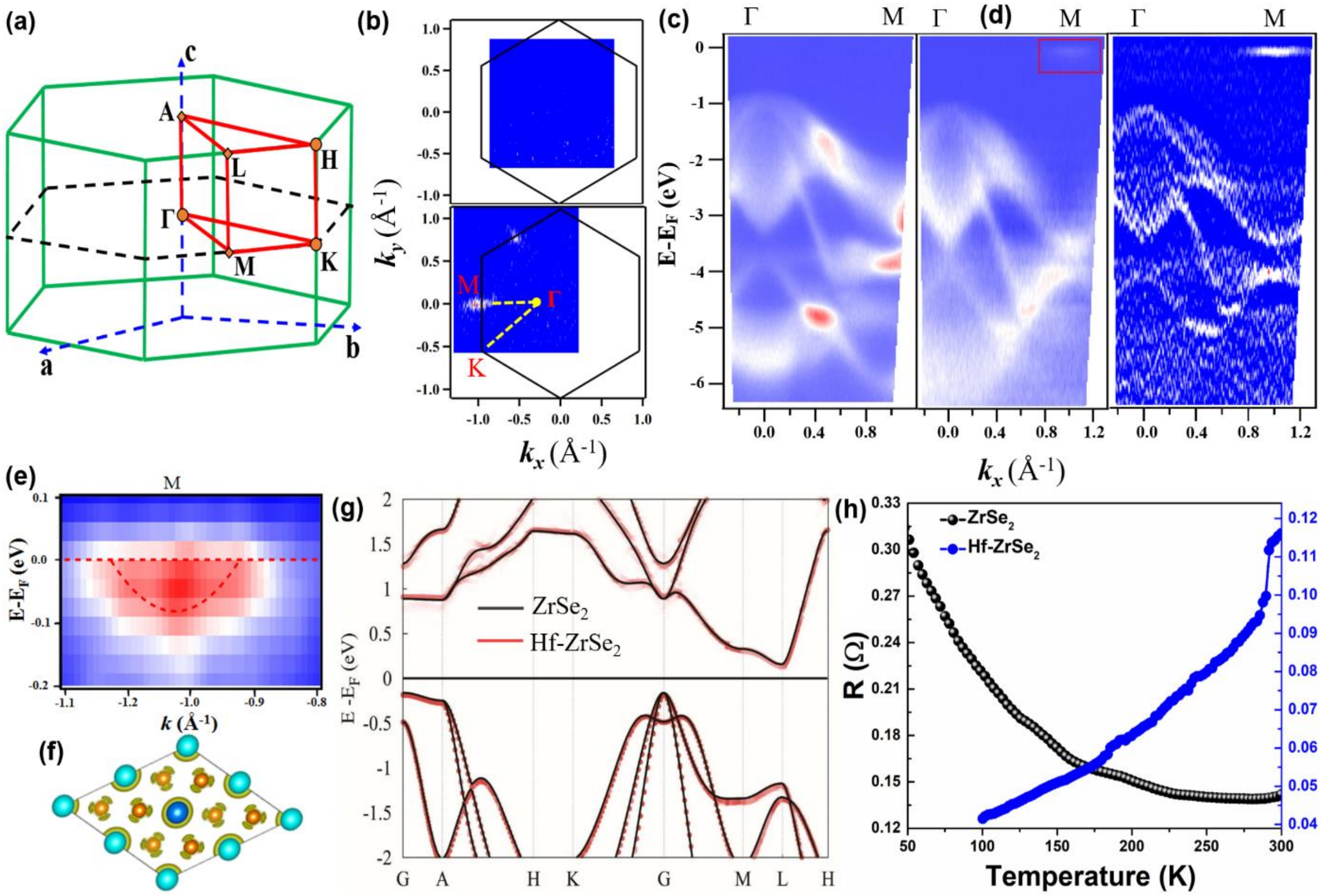

**Figure 4.** (a) The hexagonal Brillouin zone of high symmetry K-points, (b) Fermi surface plot of $ZrSe_2$ (upper panel) and $Zr_{0.88}Hf_{0.12}Se_2$ (lower panel) at $E_b = 0$ eV for high symmetry ΓM direction, are indicated by the yellow dashed lines. Band dispersion of (c) $ZrSe_2$ and (d) $Zr_{0.88}Hf_{0.12}Se_2$ and its second derivative along ΓM direction, respectively. (e) Spectral weight of the CB around M point for $Zr_{0.88}Hf_{0.12}Se_2$ as shown in (d). The data was measured at a photon energy of 33 eV and temperature of 160 K. (e) The charge density calculation and (g) the calculated unfold band

structure of $Zr_{1-x}Hf_xSe_2$ (x = 12.5%) combined with $ZrSe_2$ along the high symmetry *k*-points. (h) Resistance vs temperature (R-T) measurement of $Zr_{0.88}Hf_{0.12}Se_2$ compared with $ZrSe_2$.

**Fig. 4(e)** reveals a similar dispersion around the M point, as extracted from **Fig. 4(d)**, encircled by a red-marked square. The appeared parabolic dispersion shows the *d*-band of the Zr-*d* orbital. The CBM of $Zr_{0.88}Hf_{0.12}Se_2$ is located at 90 meV below the Fermi level, whereas the VBM is located at -1.05 eV, which indicates an indirect band gap of 0.96 eV. The results also reveal a decrease in the band gap of $ZrSe_2$ through Hf substitution rather than within those two compounds, compared to the previous work about pristine, which shows a band gap of about 1.10 eV [55]. Typically, the overall band structure remains unchanged, but merely the Hf induced more charges; due to increased extrinsic *n*-type charges in the *d* band, whose carrier density can increase the conductance [56]. In **Fig. 4(f),** we calculated the charge density of Hf-doped $ZrSe_2$. The charge density distribution suggests that around the doped atom, the distribution shows dense charge regions compared to the other atoms, which indicates that maximum charge transfer accumulation around Hf exists. Thus, Hf substitution has a strong correlation with carrier doping, leading to a change in the effective mass around the conduction. Hence, it shows the extrinsic electrons type carrier, which behaves as a metal. **Fig. 4(g)** displays the unfold band structure of $Zr_{0.88}Hf_{0.12}Se_2$ (x = 12.5%), which shows that the CB slightly shifted down toward the Fermi level and the VBM moves to lower energy as compared to $ZrSe_2$ and $HfS_2$ (see Figure S4). It can be noticed that Hf can deliver extra freedom for manipulating the electronic properties of $ZrSe_2$ due to charge carrier transfer to the Zr *d* orbital with a similar oxidation state.

Resistance vs temperature (RT) measurement was performed on $Zr_{0.88}Hf_{0.12}Se_2$ and $ZrSe_2$, as shown in **Fig. 4(h)**. The RT measurements clearly demonstrate the typical semiconducting behavior for $ZrSe_2$ with decreasing temperature resistance, while for $Zr_{0.88}Hf_{0.12}Se_2$, the resistance increases with decreasing temperature, exhibiting metallic character. Specifically, the room temperature resistance of $Zr_{0.88}Hf_{0.12}Se_2$ is lower than one-and-a-half orders of magnitude than that of $ZrSe_2$, suggesting that the conductivity of $Zr_{0.88}Hf_{0.12}Se_2$ is increased. *Taguchi* observed a similar in the transport study of $Ti_{1-x}Hf_xSe_2$ using a series of Hf substitutions after a specific limit of x value; the resistivity is decreased [57]. *Salvo* and *Waszczak* detected that the phase transition in $TiSe_2$ via substituting Ta or V donates electrons can cause a rigid band shift, which tunes their carrier densities [58]. It is suggested that the cation disorder and increase in charge density are the primary causes of this effect. *Shkvarin et al*. proposed the partial charge transfer effect from dopant to host [47] in $TiSe_2$ via Ta and V doping, which can decrease or increase the electrical conductivity of $TiSe_2$. The band below the CB is effectively induced in our sample, implying that the *d* electron of the $Hf^{4+}$ is delocalized into S the Zr *d* band, increasing the electron carrier and decreasing the hole concentration, which can affect transport properties. *Zhang et al.* [59] observed a similar trend in ZrNiSn alloy by replacing Hf on the Zr site, which can increase electron density and CB across the Fermi level.

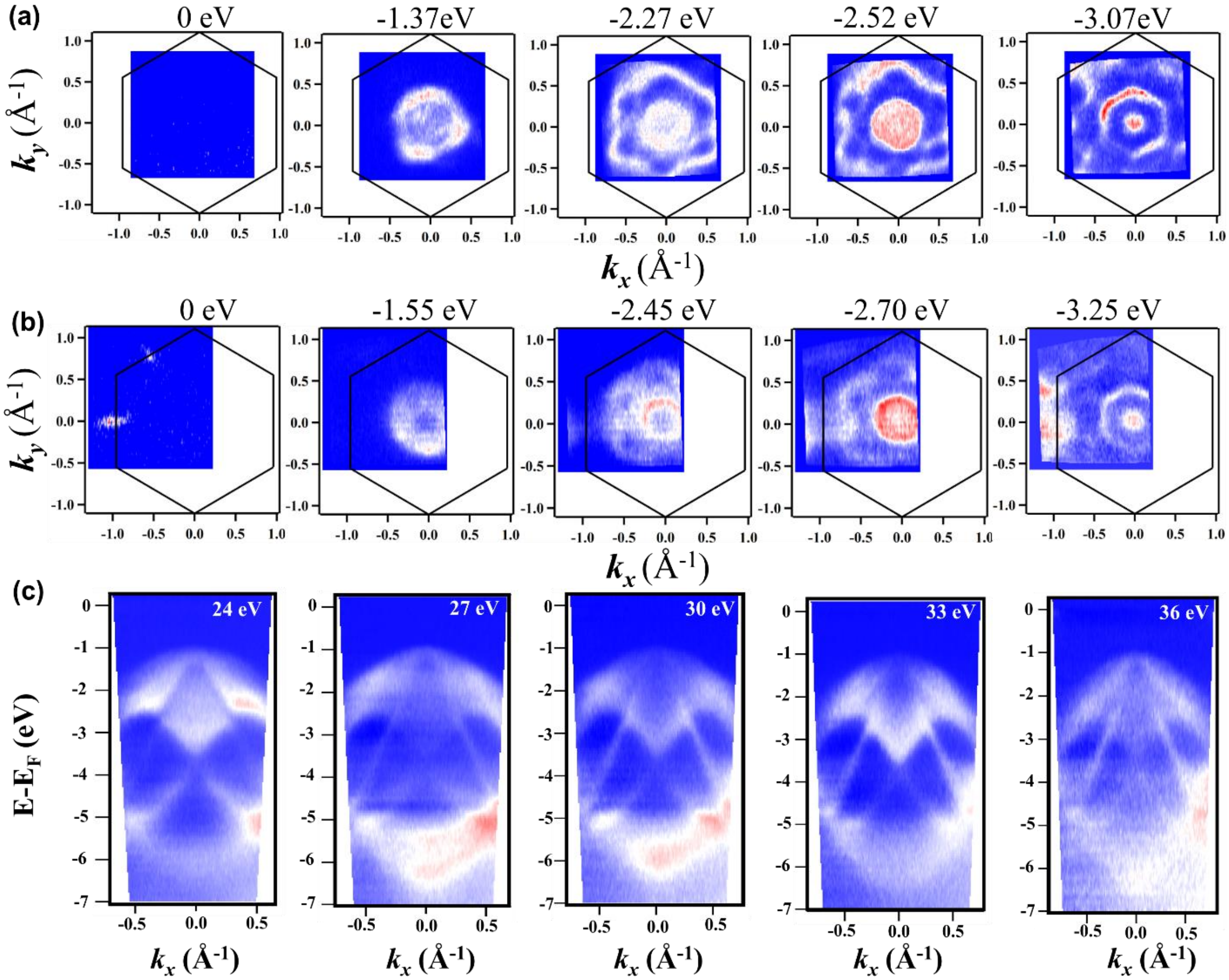

**Figure 5.** Constant energy contours of (a) pristine $ZrSe_2$ and (b) $Zr_{0.88}Hf_{0.12}Se_2$ at various binding energies, respectively. (c) Photoemission intensity was measured along Γ direction for $Zr_{0.88}Hf_{0.12}Se_2$ at 24 eV, 27 eV, 30 eV, 33 eV and 36 eV, respectively.

**Fig. 5(a & b)** shows various constant energy contours of $ZrSe_2$ and $Zr_{0.88}Hf_{0.12}Se_2$ measured within the first Brillouin zone at different binding energies. **Fig. 5(a)** shows no evident spectral weight at the Fermi surface for the parent compound at 0 eV. In contrast, **Fig. 5(b)** reveals the evident spectral weight of electron pockets at the FL around the M point in $Zr_{0.88}Hf_{0.12}Se_2$, indicating the typical electron type carriers at the FS at 0 eV. The electron pockets' spectral weight disappears at lower binding energies from the FS, indicating that these electron pockets originate from the bottom of conduction bands. In contrast, the spectral weight of the hole pocket increases for both samples, as observed in **Fig. 5(a)** and **Fig. 5(b)**, respectively. About 0.18 eV energy, difference was set between the constant energy contours for both samples, which indicates that the lattice structure of $Zr_{0.88}Hf_{0.12}Se_2$ is consistent with the parent compound, while the pocket size is a bit shifted for $Zr_{0.88}Hf_{0.12}Se_2$. This suggests that band structures move by 0.18 eV to a deeper binding energy in $Zr_{0.88}Hf_{0.12}Se_2$ due to charge doping. **Fig. 5(c)** illustrates the band dispersion plots of $Zr_{0.88}Hf_{0.12}Se_2$ at different photon energies measured along the Γ direction from 24 eV to 36 eV, respectively. While, at different photon energies, the dispersions around the VBM do not show any indistinct change, which indicates that the bands around VBM have no noticeable $k_z$ dependence, thereby preserving the overall system with other transitions. However, only at the ΓM direction can the CB at the M point cross the FL in the $Zr_{0.88}Hf_{0.12}Se_2$ compound due to the excess charge density.

As considering the excess carriers in the conduction band can be created by the disorder effects via replacing Zr with Hf, which can't be observed in similar oxidation states atoms in any other way. It can be detected that the presented disorder can be created by the excess Zr in the van der Waals gaps and other forms of lattice defects due to the substitution of the Hf. This is called a substitutional disorder. This kind of effect cannot be seen in the *k*-space if not too severe, but the energy bands can have some excess spectral weight that can be counted in the spectral functions representing, in some sense, "smeared bands". It is suggested that these extrinsic *n*-type electrons can be observed in some particular atoms, which are sensitive to these kinds of disorders, as seen in [54, 57]. Thus, band edges may become slightly prolix due to extrinsic electrons and can be extended slightly down in energy and sometimes the tail of those bands can cross the FL. This effect can reduce the effective band gap, as seen in our measurements in the form of extrinsic carriers. One might also expect that the *d* band center can appear below the FL in the band structure representing the low-energy tail of these "smeared bands", which can reach below the FL. In contrast, it could not be seen in the dispersion of the un-doped system. Therefore, it is evident that the charge density in Zr *d*-orbitals increases by extrinsic *n*-type electrons induced by the Hf atom through substitutional disorder, which can be seen from the tail of the "smeared bands" as appeared below the FL. Furthermore, the shortening of metal-metal and metal-chalcogen bond length observed from XAFS and Raman results shows evidence of extrinsic electrons in $Zr_{0.88}Hf_{0.12}Se_2$. Similar results were also observed in $V_xTi_{1-x}Se_2$ and $Ta_xTi_{1-x}Se_2$ [47, 58]. On the other hand, one might also imagine that the increased carriers may come from defects due to heavily doped atoms or self-doping. However, the data in **Fig. 2(d, h)** shows no evident defect in the sample having no super lattice points. The data suggest that the cation disorder is the major cause of this rigid band shift tail, which can cross the FL. R. D. Kirby *et al.* also predicted the effect of Zr doping in $TiSe_2$ [59]. They suggested that introducing Zr atoms could induce the impurity band or possibly result from minor deviations for exact stoichiometry with doping, showing the metallic-like effect. Therefore, the excess electrons from additional Ti presumably enter both into Γ point hole pockets and the L/M point electron pockets of Ti, which can reduce the number of holes and increase the number of electrons [59]. Those excess numbers of electrons populate the Ti *d*-band, so the conductivity at low temperatures can be changed. Thus by doping, the excess electrons are present in the $Zr_xTi_{1-x}Se_2$ doped crystals; perhaps the deviations from exact stoichiometry exist. Similarly, the recent DFT calculations showed that the excess Zr shifts the bands to lower binding energies [60].

Notably, it can be clearly seen in the band dispersion and FS contour plots that the electrons appear at the FL with Hf substitution due to excess electrons. Our study shows only the shift in the band structure and increased conductivity due to the excess electrons in the Zr *d*-band due to Hf doping. The Hf substitution on the Zr site can cause the disorder in the lattice by introducing excess Zr in the layers, which induces extrinsic electrons. The excess electrons enter the hole and electron pockets, reducing the hole number and enhancing the electron number. The excess electrons can increase the spectral weight of the conduction bands whose tail shows a rigid shift that can cross the FL. Furthermore, the VB moves to lower binding energy due to the reduction of the hole numbers. The extrinsic electrons can increase the carrier density, enhancing their conductivity and showing a metallic-like character.

## Conclusion

The electronic structure of $Zr_{0.88}Hf_{0.12}Se_2$ in compression with pristine $ZrSe_2$ single crystals synthesized by the CVT method is investigated through ARPES measurement combined with first-principles DFT calculations. We have found the weak metallic-like transition in $ZrSe_2$ with Hf substitution. ARPES measurements and DFT calculations revealed the typical semiconducting behavior of $ZrSe_2$, while $Zr_{0.88}Hf_{0.12}Se_2$ shows metallic character due to excess electrons. Interestingly, it has been observed that substituting a significant amount of Hf on the Zr site leads to creating extra Zr atoms, which can enter into the hole and electron pockets that can shift the conduction and valence band to the lower binding energy. Furthermore, the reduced bond length shows the lattice change with Hf doping. Moreover, from the ARPES results, we have observed that the overall system remains preserved, except that the conduction band's tail crossed the Fermi level due to extrinsic electrons. Similarly, the transport measurements further confirm the metallic character of $Zr_{0.88}Hf_{0.12}Se_2$ due to excess charge carriers. These results highlight the importance of cation substitution-induced transition in 2D layered materials, which can enable these materials for future quantum electronics and valleytronic applications.

## Acknowledgements

This research was supported by the by Princess Nourah bint Abdulrahman University Researchers Supporting Project number (PNURSP2024R1), Princess Nourah bint Abdulrahman University, Riyadh, Saudi Arabia, National Natural Science Foundation of China (Grants No. 62150410438), the Beihang Hefei Innovation Research Institute (project no. BHKX-19-02) and Deanship of Scientific Research at King Khalid University, Saudi Arabia, under grant number RGP. 2/445/44. We thank the Beijing Synchrotron Radiation Facility (1W1B, BSRF) for the XAFS experiments. We acknowledged beamline 13U of the National Synchrotron Radiation Laboratory (NSRL) for the ARPES experiments.